\documentclass[journal]{IEEEtran}
 
\usepackage{graphicx}
\usepackage{float}
\usepackage[utf8]{inputenc} 
\usepackage[T1]{fontenc}    
\usepackage{hyperref}       
\usepackage{url}            
\usepackage{booktabs}       
\usepackage{amsfonts}       
\usepackage{amsmath} 
\DeclareMathOperator*{\argmax}{argmax} 
\DeclareMathOperator*{\argmin}{argmin} 
\usepackage{lipsum}
\usepackage{verbatim}
\usepackage{makecell}
\usepackage{multirow}
\usepackage{bm}
\usepackage{url}
\usepackage{xcolor,soul}
\usepackage[most]{tcolorbox}
\newtcolorbox{highlighted}{colback=yellow,coltext=black,breakable}
\usepackage{caption}
\ifCLASSINFOpdf
  \else
\fi

\begin{document}

\title{Active Learning for Sound Event Detection}

\author{Zhao~Shuyang,
        Toni~Heittola,
        and~Tuomas~Virtanen
        \thanks{The research leading to these results has received funding from the European Research Council under the European Unions H2020 Framework Programme through ERC Grant Agreement 637422 EVERYSOUND. }

}

\maketitle

\begin{abstract}
This paper proposes an active learning system for sound event detection (SED). It aims at maximizing the accuracy of a learned SED model with limited annotation effort. The proposed system analyzes an initially unlabeled audio dataset, from which it selects sound segments for manual annotation. The candidate segments are generated based on a proposed change point detection approach, and the selection is based on the principle of mismatch-first farthest-traversal. During the training of SED models, recordings are used as training inputs, preserving the long-term context for annotated segments. The proposed system clearly outperforms reference methods in the two datasets used for evaluation (TUT Rare Sound 2017 and TAU Spatial Sound 2019). Training with recordings as context outperforms training with only annotated segments. Mismatch-first farthest-traversal outperforms reference sample selection methods based on random sampling and uncertainty sampling. Remarkably, the required annotation effort can be greatly reduced on the dataset where target sound events are rare: by annotating only 2\% of the training data, the achieved SED performance is similar to annotating all the training data.

\end{abstract}


\IEEEpeerreviewmaketitle

\section{Introduction}
Sound event detection (SED) is a task of automatically identifying sound events such as gunshot, glass smash, and baby cry from an audio signal. It predicts the presence of each target sound event and its onset/offset. SED has been applied in various applications, including noise monitoring~\cite{Maijala2018_APACOUST}, healthcare monitoring \cite{DBLP:journals/jcse/GoetzeSGHAW12}, wildlife monitoring~\cite{DBLP:journals/corr/abs-1905-08352}, urban analysis \cite{DBLP:conf/eusipco/SalamonB15}, and multimedia indexing and retrieval \cite{DBLP:journals/taslp/CaiLHZC06}.

Due to the large number and variability of sound events in real-life acoustic environments, there does not exist a universal SED model. Most SED applications require their own models. The development of a SED model is commonly based on supervised learning, which typically requires a large amount of labeled data as training material. Compared to capturing audio, annotating them is much more time-consuming in most cases. Thus, a practical problem is to optimize the SED accuracy with a limited annotation effort.

Recently, weakly supervised learning has been studied to reduce the required annotation effort in the development of SED models \cite{DBLP:journals/taslp/KongYXIWP19, DBLP:journals/taslp/McFeeSB18}. Weak labels indicate the presence of target event classes in an audio signal, without temporally locating them. In most cases, assigning weak labels is much simpler, compared to assigning strong labels, which requires the onset/offset of each individual sound event. 

Despite the existence of weakly supervised learning, annotating a large amount of data is still time-consuming. 
Active learning has been used in various machine learning problems \cite{DBLP:conf/www/HoiJL06, DBLP:journals/jcisd/Liu04a}, where labels are difficult, time-consuming, or expensive to obtain. 
An active learning algorithm controls a labeling process by selecting the data to be labeled, typically based on an estimate of the capability to improve an existing model. In most cases, active learning targets the situation where unlabeled data is abundant, but the amount of annotations that can be made is limited. The total duration of audio that can be manually labeled is called a labeling budget.

Active learning for SED has previously not been studied, though a few active learning studies have been made on sound classification \cite{DBLP:journals/taslp/RiccardiH05, DBLP:conf/icassp/Hakkani-TurRG02, 10.1371/journal.pone.0162075, DBLP:conf/icassp/ZhaoHV17, DBLP:conf/iwaenc/ZhaoHV18}. All of these studies are limited to single-label classification on sound segment datasets \cite{DBLP:conf/mm/Piczak15, DBLP:conf/mm/SalamonJB14}, where a sound segment contains an isolated event. However, the situation is different in SED, which typically deals with long signals containing many sound events, possibly overlapping in time. In this paper, we propose an active learning system for SED. The proposed system includes the following novelties: (i) Variable-length sound segments are generated as selection candidates using a change point detection approach. To the best of our knowledge, audio change point detection has previously not been used for active learning. Change point detection is used to avoid generating segments that contain only a part of an event, which is sometimes hard to recognize either manually or automatically. (ii) The selection of candidate segments is based on the mismatch-first farthest-traversal principle, which has been shown effective in sound classification \cite{DBLP:conf/iwaenc/ZhaoHV18}. In this study, the selection principle is generalized to the whole labeling process, without clustering in the first stage as is originally proposed. As a result, the process does not require the cluster number as a hyper-parameter, which is sometimes hard to estimate. Furthermore, the sample selection method is extended to multi-label classification. (iii) We propose to use a partial sequence loss during the training of SED models, to preserve the temporal context of annotated segments: each recording is used as training input and the training loss is computed based on only the outputs within annotated segments. Previously, segments generated from the same recordings are processed independently in the training, such as in UrbanSound8K~\cite{DBLP:conf/mm/SalamonJB14} and AudioSet~\cite{DBLP:conf/icassp/GemmekeEFJLMPR17}. 

The structure of the rest of the paper is as follows. Related works are discussed in Section~II. The proposed system is introduced in Section~III. The evaluation of the proposed system is presented in Section~IV. The conclusions are drawn in Section~V. 

\section{Related works}
 
\subsection{Weakly supervised learning}
Weakly supervised learning has recently attracted lots of research interests in the field of SED, especially after the release of a large publicly available sound event dataset, AudioSet \cite{DBLP:conf/icassp/GemmekeEFJLMPR17}, which provides only weak labels. AudioSet has been used to learn high-level representations in  \cite{DBLP:conf/icassp/KumarKF18}. The learned representation clearly outperforms hand-crafted features such as log-mel spectrogram in an environmental sound classification dataset \cite{DBLP:conf/mm/Piczak15} and an acoustic scene classification dataset \cite{DBLP:conf/eusipco/MesarosHV16}. Furthermore, weakly supervised learning can be also used to directly learn SED models, such as in Detection and Classification of Acoustic Scenes and Events (DCASE) 2017 task 4 subtask B \cite{DCASE2017challenge}.

Previous weakly supervised learning studies ~\cite{DBLP:journals/taslp/KongYXIWP19, DBLP:journals/taslp/McFeeSB18, DBLP:conf/icassp/KumarKF18,DBLP:conf/icassp/0004KWP18,DBLP:conf/icassp/0005LM19} use pooling functions to aggregate frame-level class probabilities into segment-level. Among the studied pooling functions, attention pooling \cite{DBLP:journals/taslp/KongYXIWP19, DBLP:conf/icassp/0004KWP18} appears to be the most popular one \cite{DBLP:conf/icassp/0005LM19}.  Besides the class probability, an attention neural network \cite{DBLP:journals/taslp/KongYXIWP19, DBLP:conf/icassp/0004KWP18} predicts a pooling weight for each frame. The segment-level output is based on the weighted average of the frame-level class probabilities. Besides attention pooling, softmax pooling has also been shown effective in \cite{DBLP:journals/taslp/McFeeSB18, DBLP:conf/icassp/0005LM19}. An adaptive pooling method, introducing a learnable hyper-parameter to softmax pooling  \cite{DBLP:journals/taslp/McFeeSB18},  achieved similar SED performances compared to using strong labels, in three SED datasets.

 \subsection{Sample selection}
There are different problem setups defined in the field of active learning. Previous studies on sound classification follow the setup of pool-based sampling, where a large collection of unlabeled data is available from the very beginning of a labeling process. 
Uncertainty sampling method was studied in \cite{DBLP:journals/taslp/RiccardiH05, DBLP:conf/icassp/Hakkani-TurRG02, 10.1371/journal.pone.0162075}, where the uncertainty to classify a sample with an existing model was used for sample selection. One of the problems with uncertainty sampling is the unreliable certainty estimation unless a decent amount of data is labeled. In many cases, uncertainty sampling does not outperform random sampling when the labeling budget is low \cite{DBLP:journals/taslp/RiccardiH05, DBLP:conf/icassp/Hakkani-TurRG02}. Another problem with uncertainty sampling is the low diversity in a selection batch, since the samples uncertain to the same model are often similar \cite{DBLP:conf/www/HoiJL06, DBLP:conf/acl/Sassano02}.

Cluster-based active learning was proposed in \cite{DBLP:conf/icassp/ZhaoHV17}. Segment-to-segment similarities were measured based on the distribution of MFCCs in each sound segment in the training dataset. $K$-medoids clustering was performed on the sound segments, and the centroids of clusters (medoids) were selected for annotation. The method is called medoid-based active learning (MAL). A label assigned to a medoid was propagated to all segments within the same cluster. When all the medoids were annotated, another round of clustering was performed. Both the annotated labels and the propagated labels were used in training acoustic models. MAL relies completely on the similarity measurement. The advantage is that it enables good performance with a low labeling budget, since it does not require a reliable model. However, the method is not optimal as the labeling budget grows, since the selection of samples does not take previously annotated samples into account. Another problem is that the choice of the number of clusters $K$ requires a prior knowledge about a dataset.

As an extension of MAL, mismatch-first farthest-traversal was proposed in \cite{DBLP:conf/iwaenc/ZhaoHV18}. It performs only one round of $K$-medoids clustering as the first stage. After annotating the medoids, the sample selection is continued with mismatch-first farthest-traversal as the second stage. The samples with mismatched predictions were selected as the primary criterion. They were further selected by their distances to previously selected samples as the secondary criterion. The target is to maximize the diversity of selected samples. The first stage of the method is equivalent to MAL, and the second stage, which starts at the labeling budget of $k$, clearly outperforms the original MAL and other reference methods with all evaluated labeling budget. In addition, an approach was proposed to estimate the cluster number $K$. However, it assumed a relatively balanced number of instances from each sound class. This assumption can hardly be satisfied in SED problems.

In comparison to the previous active learning studies on sound classification \cite{10.1371/journal.pone.0162075, DBLP:conf/icassp/ZhaoHV17, DBLP:conf/iwaenc/ZhaoHV18}, the problem setup in this study has the following differences. Firstly, generating segments for annotation is considered as a part of the active learning system in this study, whereas previous studies utilize sound segments that are already generated before the active learning process. Secondly, this study allows a set of classes assigned to a segment, whereas the previous studies require exactly only one class assigned to a segment. Thirdly, this study predicts not only the event class as the previous studies, but also the onset and offset of each individual event.

\section{The proposed method}

\begin{figure}[t]
  \centering
  \includegraphics[trim={20mm 30mm 20mm 20mm},clip,width=0.42\textwidth]{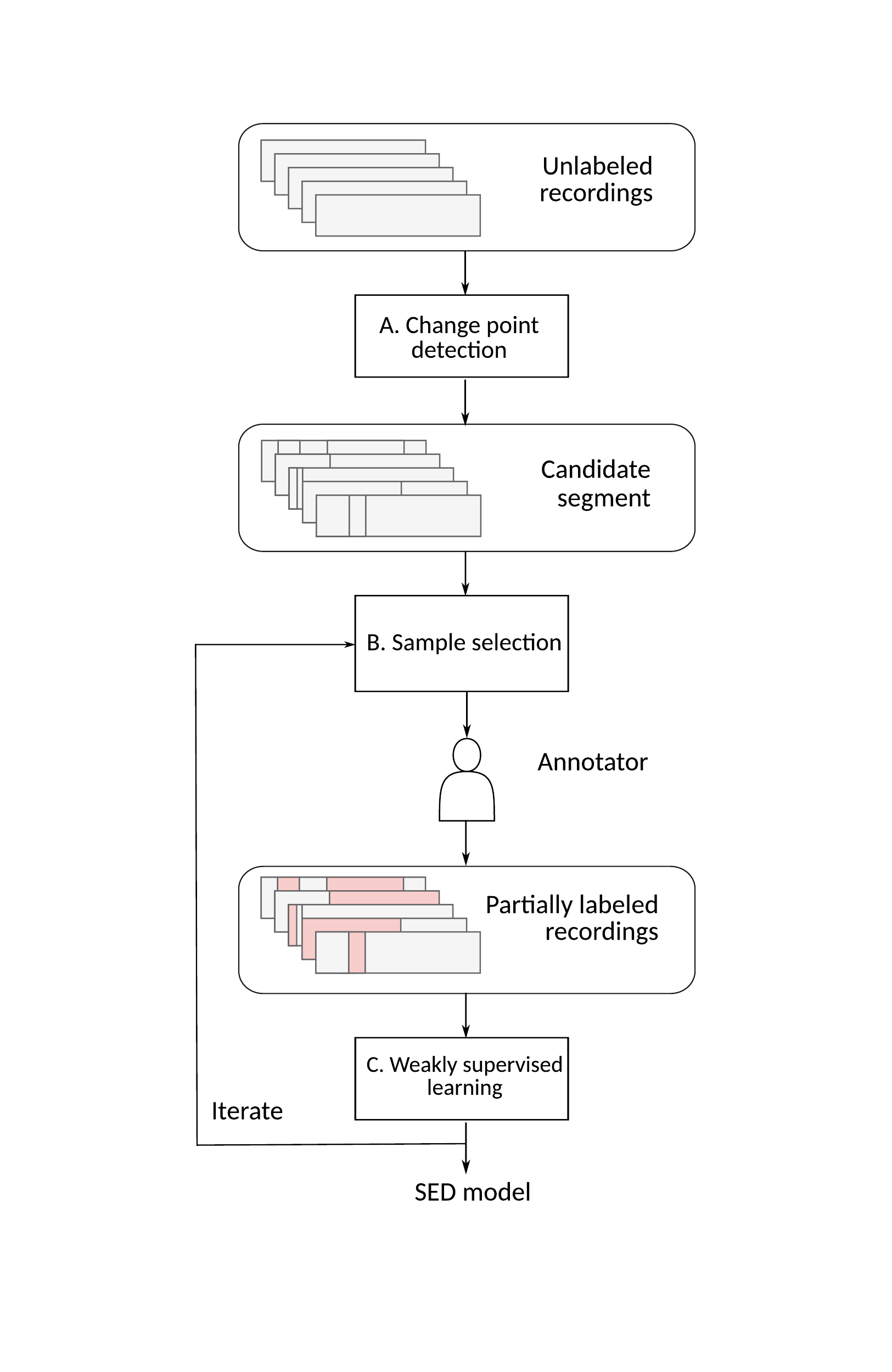}
  \caption{The overview of the proposed active learning system. The three processing blocks correspond to the three subsections in Section II.}
  \label{fig:flowchart1}
\end{figure}

 The proposed active learning system aims at optimizing the  accuracy of a learned SED model, with a limited annotation effort. The general overview of the proposed system is illustrated in Figure~\ref{fig:flowchart1}. It takes a set of unlabeled audio recordings as input and outputs a SED model. A human annotator is required to assign labels to sound segments that the system selects from the recordings. The SED model is trained with annotated sound segments.

 At the beginning of the active learning process, change point detection is performed, splitting each recording into segments. Each segment, later called a sample, is used as a candidate for being selected to be annotated. The definition of sample, sampling, and training example follows [13]. The active learning process is iterative, following batch mode active learning scheme [8]. In each iteration, a batch of samples is selected for annotation, and a SED model is trained with annotated samples. The sample selection is based on mismatch-first farthest-traversal. Mismatch-first as the primary criterion targets on the samples that are previously wrongly predicted. Farthest-traversal as the secondary criterion aims at maximizing the diversity of selected samples. 

In order to save annotation effort, the system requires only weak labels that are assigned to individual segments. In each recording, the annotated segments are visualized in pink in Fig.1. During the training of SED models, original recordings are used as training inputs, regarded as partially labeled sequences. The training loss is derived from only the annotated parts of each recording, and the unlabeled parts are used to provide context information.

\subsection{Change point detection}
\begin{figure}[t]
  \centering
  \includegraphics[trim={0mm 170mm 20mm 0mm},clip,width=0.49\textwidth]{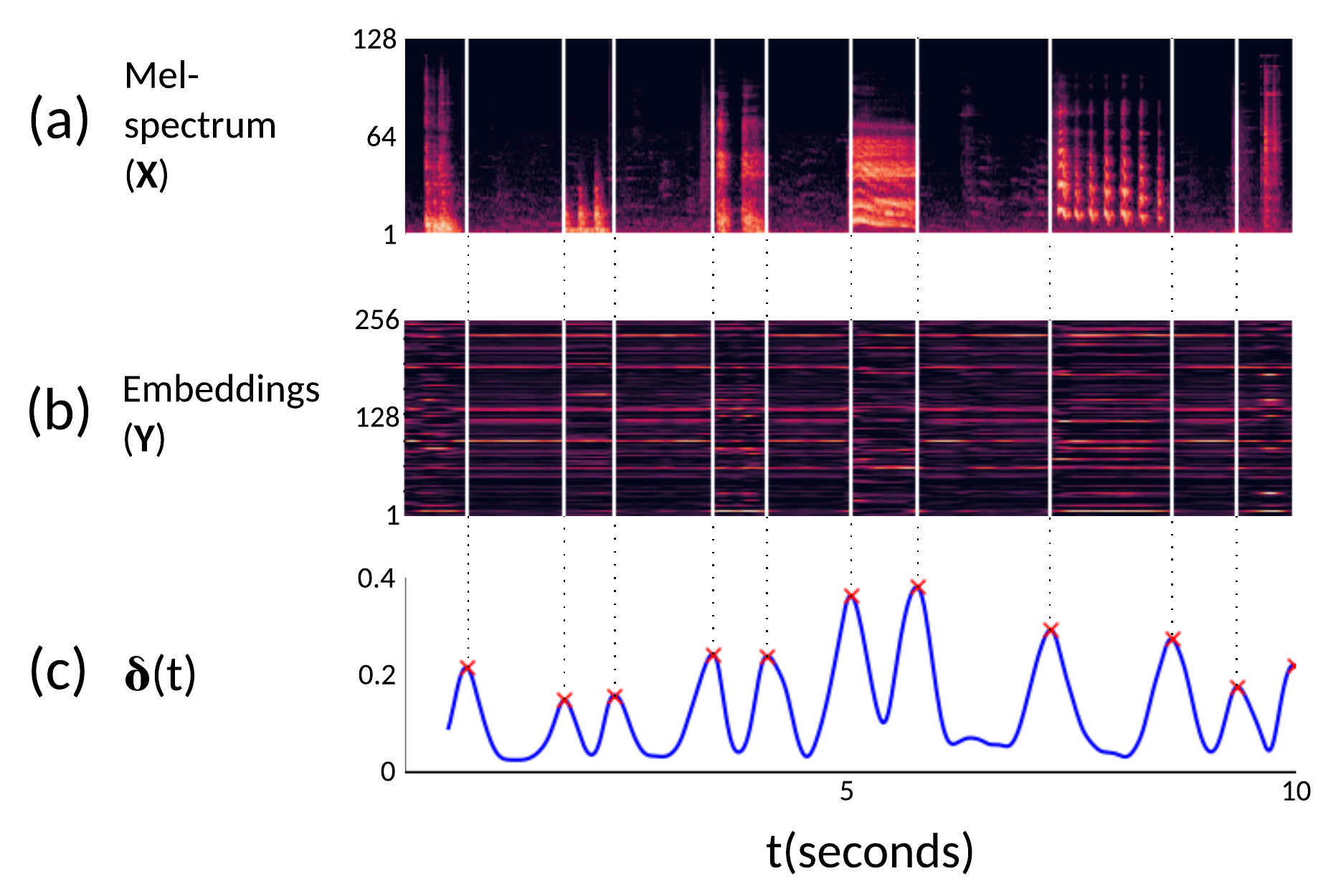}
  \caption{Panel (a) is the log-mel spectrogram of an example audio signal, with the detected change points marked by white vertical lines. Panel (b) visualizes the embeddings extracted using a pre-trained model. Panel (c) illustrates the estimated likelihood of change on each time step. The peaks in the likelihood sequence are detected as change points, which are marked with red crosses.}
  \label{fig:seg}
\end{figure}

In the proposed system, recordings are first split into short segments, as illustrated in Figure~\ref{fig:seg}. Short segments have two advantages over full recordings as basic units for annotation. The temporal resolution of weak labels, indicating event presences in each recording, is sometimes insufficient to train SED models, especially when sound events are dense. In addition, the diversity of acoustic content in a recording is sometimes limited, since the sounds are typically produced from the same sources. In many cases, annotating only representative segments within each recording is sufficient.

The segments are generated based on a change point detection approach, in order to obtain segments containing complete sound events, since segments with only part of an event are sometimes difficult to annotate. Aiming at discriminative features for sound event activities, embeddings are extracted per frame using a pre-trained model. The architecture of the pre-trained model follows the network architecture used in [21].  The details of the architecture is described in Section III C. The training material and validation criterion used for training the pre-trained model generally follows the setup in [18]. Change point detection is performed on the embeddings $\mathbf{Y} = [\mathbf{y}_1, ..., \mathbf{y}_T]$, where each embedding vector $\mathbf{y}_t$ corresponds to the time frame $t=1,2,...T$. A likelihood of a change $\delta(t)$ is measured for each time frame $t$ by the cosine distance between the means of the past $M$ frames and the future $M$ frames. The $M$ frames correspond to 0.5 seconds, thus one second is the length of the analysis window for the estimation of $\delta(t)$. Previous unsupervised audio segmentation approaches are mostly proposed for speaker diarization
\cite{DBLP:journals/taslp/KottiBK08, DBLP:conf/icassp/ChengWF08}. These methods typically use a fixed or variable length analysis window around two seconds, based on the expected duration of speaker utterances \cite{DBLP:journals/taslp/KottiBK08}. This study uses an analysis window of one second based on the expected duration of short sound events such as gunshot or glass break.

The panel (c) in Figure~\ref{fig:seg} illustrates the likelihood of change estimated at each frame in an example audio signal. A peak in the likelihood is used as a change point. The change points divide an audio signal into segments, which are used as candidates for sample selection and annotation.

\subsection{Sample Selection}
\begin{figure}[t]
  \centering
  \includegraphics[width=0.48\textwidth]{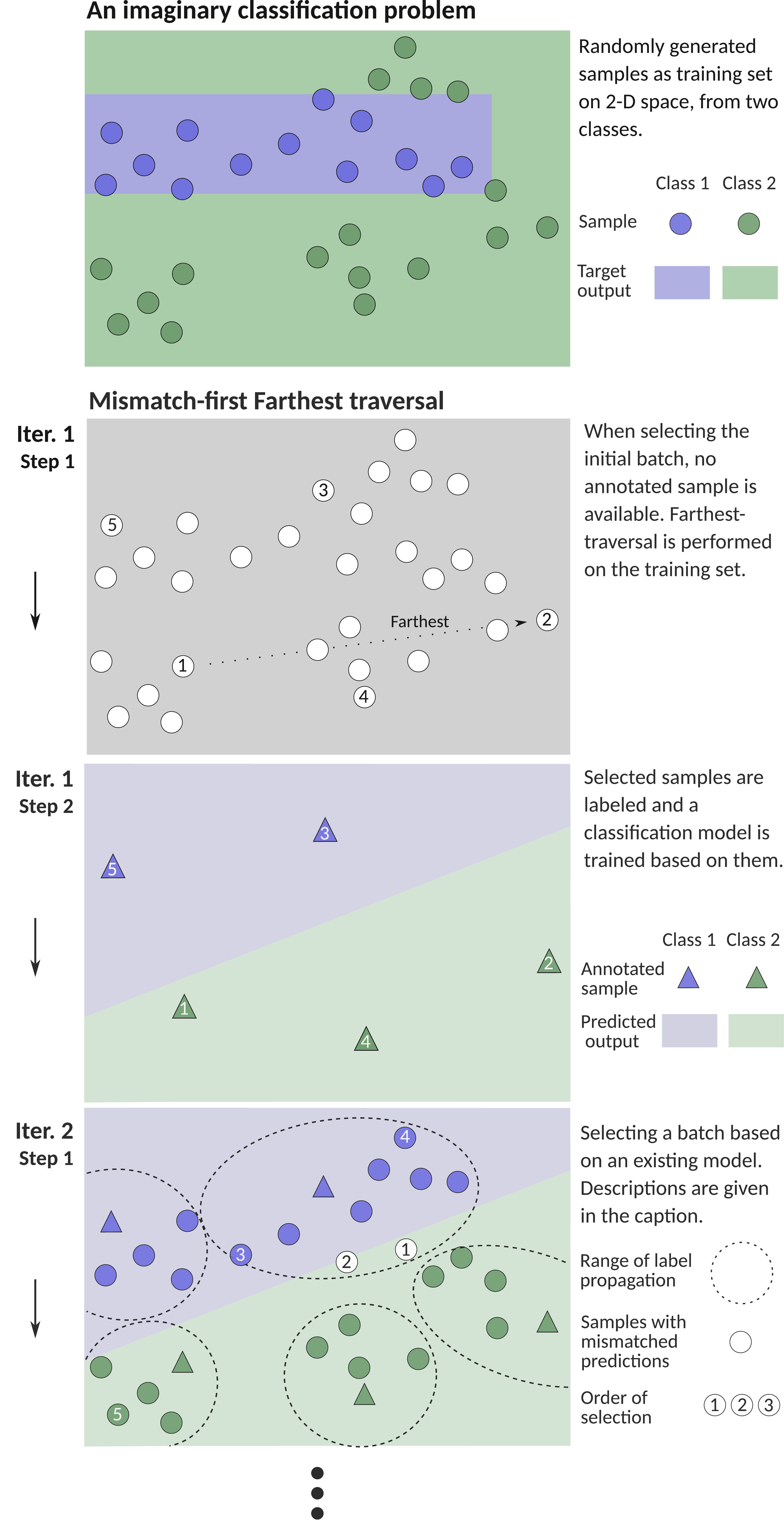}
  \caption{A visualization of mismatch-first farthest-traversal on an imaginary binary classification problem. In the bottom panel, the range of label propagation is used to visualize the area where an annotated data point propagates its label. Farthest-traversal is first performed on samples where propagated labels mismatch with model predictions, and then on samples with matched predictions.}
  \label{fig:flowchart2}
\end{figure}

Figure~\ref{fig:flowchart2} illustrates the active learning process with the generated candidate segments as samples. The sample selection method follows the principle of mismatch-first farthest-traversal \cite{DBLP:conf/iwaenc/ZhaoHV18}. Detailed visualization of the sample selection method is given online\footnote{\url{https://github.com/zhao-shuyang/active_learning}}. 

When selecting the first batch of samples, no annotated samples are available. In order to maximize the diversity of selected samples, farthest-traversal is performed on the whole training set. Farthest-traversal is explained later in this section. An annotator assigns labels to the selected samples, with which a SED model is trained. 

Two types of predicted labels are generated for each unlabeled sample. Based on a trained SED model, \textit{model-predicted labels} are generated. Based on the nearest neighbor prediction, \textit{propagated labels} are generated, according to a distance metric. The similarity between the two types of predicted labels is measured for each unlabeled sample. The measurement of the prediction similarity is given in the subsection about the mismatch-first criterion. The samples are primarily ranked by the prediction similarities, lowest first. There are typically multiple samples with the same prediction similarities. They are further ranked by the distance to the previously selected samples, farthest first. A batch of samples with the highest rank is presented to the annotator and the active learning process continues to the next iteration.

\subsubsection{Mismatch-first criterion}

\begin{figure}[t]
  \centering
  \includegraphics[trim={0mm 200mm 2mm 0mm},clip,width=0.48\textwidth]{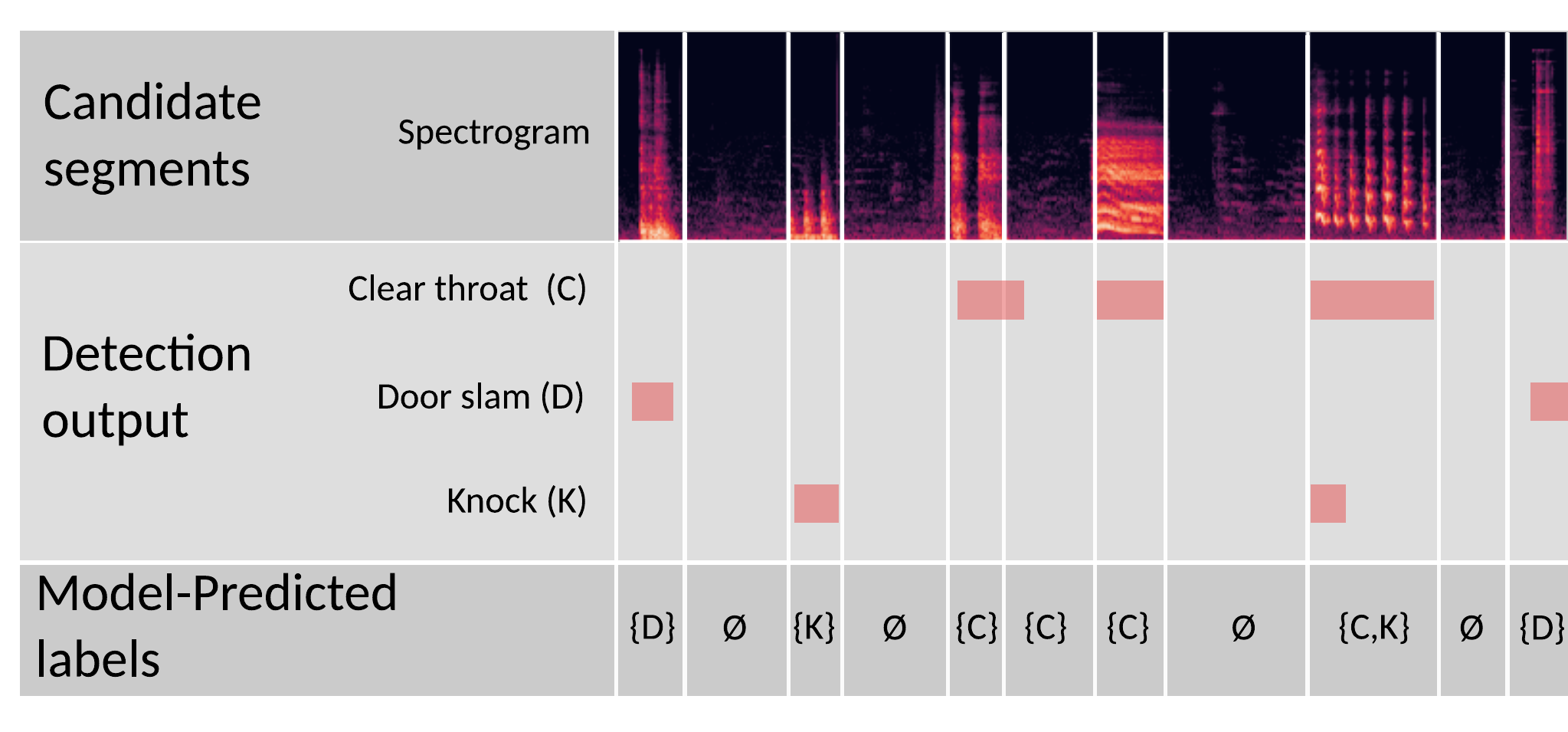}
  \caption{An example of deriving model-predicted labels from sound event detection output.}
  \label{fig:model-prediction}
\end{figure}

At the beginning of each iteration, except the first one, model-predicted labels and propagated labels are generated for each unlabeled sample. Model-predicted labels are derived from the SED outputs of each recording as is illustrated in Figure~\ref{fig:model-prediction}. When a class of sound event is detected within a candidate segment, a model-predicted label is generated, associating the class of the sound event to the segment. The classes associated with a sample $x$ according to the SED outputs are denoted as a set $\mathcal{A}_x$. Propagated labels are generated based on the nearest neighbor prediction. Each unlabeled sample $x$ is assigned the labels of its nearest annotated sample. The distance between two samples is measured by the cosine distance between the means of embeddings within the two samples. These propagated labels are denoted as a set $\mathcal{B}_x$. 

In a multi-label classification problem, the similarity between the propagated labels and the model-predicted labels on a sample $x$ is measured based on the Jaccard index as, 

\begin{equation}
    J(x) = 
    \begin{cases}
    \frac{|\mathcal{A}_x \cap \mathcal{B}_x|}{|\mathcal{A}_x \cup \mathcal{B}_x|} &, \text{if  }  \mathcal{A}_x \cup \mathcal{B}_x \neq \emptyset \\
    1 &, \text{if  }  \mathcal{A}_x \cup \mathcal{B}_x = \emptyset
    \end{cases}.
\end{equation}

Samples are first selected within the set $\mathcal{M}$, which consists of the samples with the lowest prediction similarities among the set of unlabeled samples. 

The mismatch-first criterion is based on an assumption that a model benefits more from a counterexample, where it makes an error, in comparison to an example where it makes a correct prediction. When the prediction results based on two mechanisms mismatch, the sample is a counter example for at least one of the mechanisms. Since the nearest neighbor prediction and neural network prediction are two fundamentally different mechanisms, their prediction results are usually supplementary information to each other. In addition, the two prediction mechanisms are based on different contexts. The nearest neighbor prediction is based only on annotated segments, whereas the SED model uses original recordings as a context for annotated segments.

\subsubsection{Farthest-traversal}
Farthest-traversal aims at optimizing the diversity of selected samples. It selects the sample farthest to the previously selected samples. The distance between two samples is measured by the cosine distance between the means of embeddings within the two samples. The previously selected samples are denoted as a set~$\mathcal{S}$, which is the union of annotated samples and the samples already selected in the current iteration. As a result, a selected sample is neither similar to annotated samples, nor to the ones to be annotated in the same batch. The distance from a sample $x$ to the set of previously selected samples $\mathcal{S}$ is defined as $d(x, \mathcal{S}) = min_{y \in \mathcal{S}} d(x,y)$. 

 With mismatch-first as the primary criterion and farthest-traversal as the secondary criterion, a sample is selected as 
\begin{equation}
    s = \argmax_{x \in \mathcal{M}}{d(x, \mathcal{S})}, 
\end{equation}
where $\mathcal{M}$ is the set of samples with the lowest prediction similarities.

The selected samples are added one by one into a selection batch and removed from the set of unlabeled samples until the batch reaches a pre-defined batch size. After that, the batch of selected samples is presented to the annotator, querying for weak labels. Weak labels of a segment is a set of sound event classes, that are present in the segment. 

Previous active learning studies on sound classification incorporate the idea of semi-supervised learning, where predicted labels on unlabeled data are also used in training \cite{10.1371/journal.pone.0162075, DBLP:conf/icassp/ZhaoHV17, DBLP:conf/iwaenc/ZhaoHV18}. Since semi-supervised learning techniques have been rapidly developed in recent years, this study considers semi-supervised learning as a separate problem and focuses only on active learning. The optimal combination with semi-supervised learning is considered as future work.

\subsection{Weakly supervised learning}

\begin{figure}[t]
  \centering
  \includegraphics[width=0.49\textwidth]{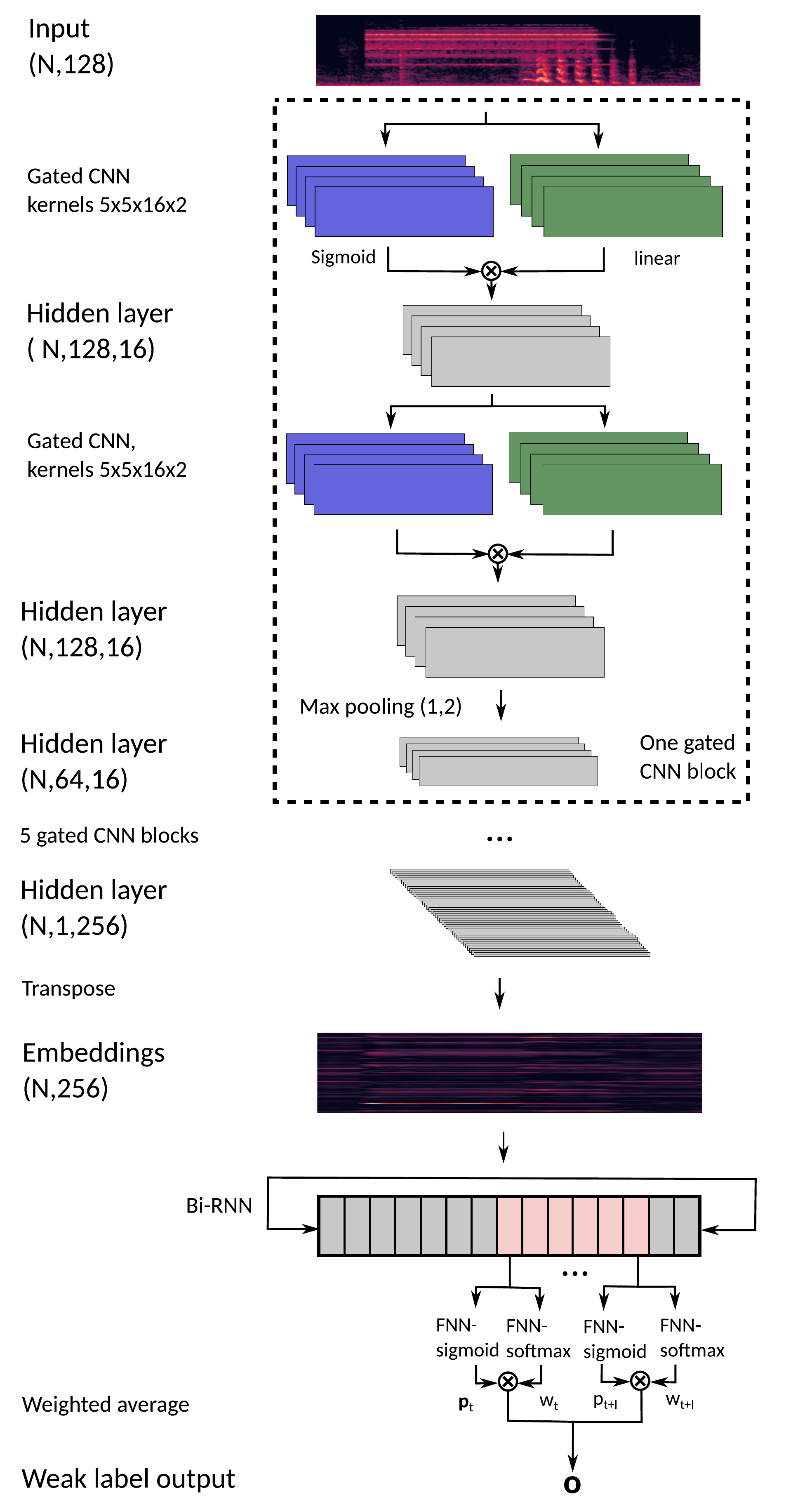}
  \caption{The diagram of the network architecture used in weakly supervised learning. The frames marked red in the bidirectional RNN outputs correspond to an annotated segment.}
  \label{fig:network}
\end{figure}

Previous active learning studies  \cite{10.1371/journal.pone.0162075, DBLP:conf/icassp/ZhaoHV17, DBLP:conf/iwaenc/ZhaoHV18} use support vector machine to classify sound segments. This study uses a neural network to perform SED, since neural networks are commonly used for SED problems. The architecture of the network follows an attention-based weakly supervised learning system \cite{DBLP:conf/icassp/0004KWP18}, which ranks the 1st in the audio tagging subtask and the 2nd in SED subtask in a weakly supervised learning challenge, DCASE 2017 task 4. In \cite{DBLP:conf/icassp/0004KWP18}, each training input is an annotated segment sliced from a YouTube video. In comparison, this study uses each original recording as a training input, preserving the context for annotated segments.

 The network architecture is illustrated in Figure~\ref{fig:network}. The input of the network is the log-mel spectrogram of a recording, denoted as $\mathbf{X}=[\mathbf{x}_1, ..., \mathbf{x}_T]$, where each vector $\mathbf{x}_t$ represents the log-mel band energies in a time frame $t = 1,2,...T$. The target output is a vector $\bm{\tau}$, corresponding to the event class activities. Each element in the target output vector $\bm{\tau} = [\tau_1, ..., \tau_C]$ represents the presence/absence of an event class, 0 for absence and 1 for presence, and $C$ denotes the number of classes. 
 
 The network consists of six blocks of gated CNNs, each of which consists of a linear CNN layer and a sigmoid CNN layer. The element-wise product between the outputs of the two CNN layers is fed to the next layer. Compared to traditional CNNs that use rectified linear units as activation function, the gated CNNs reduce the gradient vanishing problem in a deep structure \cite{DBLP:conf/icml/DauphinFAG17}. The gated CNNs transfer the input log-mel spectrogram into a sequence of embeddings $\mathbf{Y} = [\mathbf{y}_1, ..., \mathbf{y}_T]$, where an embedding vector $\mathbf{y}_t$ corresponds to a time step $t$. In order to model a long-term temporal context, three bi-directional gated recurrent unit (GRU) layers are used. The GRUs process the embedding sequence, and output a vector $\mathbf{y}'_t$ in each time step. A fully-connected sigmoid layer is used to estimate the class probabilities in each time step as $\mathbf{p}_t = cla( \mathbf{y}'_t)$. In parallel, a fully-connected softmax layer estimates the pooling weights as $\mathbf{w}_t = att(\mathbf{y}'_t)$. 
 
  In order to derive the output for an annotated segment, the weighted average of the class probabilities is computed across all frames within the segment. Given the start time point of a segment as $t$ and the length of it as $l$, the weak label output of the segment is computed as

\begin{equation}
    \mathbf{o} = \frac{\sum_{i=t}^{t+l}{\mathbf{w}_i \cdot \mathbf{p}_i}}{\sum_{j=t}^{t+l}{\mathbf{w}_j}},
\end{equation}
where $\cdot$ represents element-wise multiplication. Binary cross-entropy is used to measure the loss between the prediction output $\mathbf{o}$ and the target output $\bm{\tau}$ for each annotated segment, as 

\begin{equation}
L(\bm{\tau}, \mathbf{o}) = \sum_{k=1}^C-{(o_k\log(\tau_k) + (1 - o_k)\log(1 - \tau_k))},
\end{equation}

where $C$ is the number of classes. The training loss for a recording is the sum of the loss from each annotated segment within it.

In this study, the gated CNNs that extract embeddings are pre-trained with the balanced set of AudioSet \cite{DBLP:conf/icassp/GemmekeEFJLMPR17}. The embedding extraction function is considered as a general knowledge, which can be transferred to different SED problems. During the pre-training, the GRU layers are not used, and embedding vectors are directly fed to the fully-connected layers. The output of the second last layer of a classification network is used as embeddings. This follows the common practice in previous transfer learning studies \cite{DBLP:conf/icassp/KumarKF18, DBLP:conf/nips/AytarVT16} on sound classification.

In the active learning process, the pre-trained embedding extraction function $e$ is fixed. The parameters of the GRU layers $gru$, the sigmoid layer $cla$, and the softmax layer $att$ are trained with data annotated in the active learning process. With a limited labeling budget, usually a small number of segments are labeled in each recording. During the training, the log-mel spectrogram of full recordings are used as input, but the training loss is derived from only the frames corresponding to labeled segments. When performing SED on test data, the detection output is based on the class probabilities, the output of $cla$, without using the layer $att$.

Previous studies \cite{DBLP:conf/mm/SalamonJB14,DBLP:conf/icassp/GemmekeEFJLMPR17, DBLP:conf/icassp/ZhaoHV17, DBLP:conf/iwaenc/ZhaoHV18} use each annotated segment as input, instead of the original recordings. As a result, they lose the contextual information in the original recordings. The contextual information may benefit the SED performance from different aspects. Firstly, given background sounds as contextual information, a model can learn the unique characteristics of an event out of the background. Secondly, the contextual information can be used to model the dependencies between acoustic events and scenes. For example, it is common to hear key rattling before door opening and it is common to hear a bird chirping in a forest.

\section{Evaluation}
In order to evaluate the performance of the proposed system, two sets of experiments are made on two different datasets. The first one focuses on the training input and annotation unit. The second one focuses on the sample selection method. 

\subsection{Datasets and settings}

In order to evaluate active learning performances with different SED scenarios, two SED datasets are used in the evaluation. The statistics comparing the two datasets are shown in Table~\ref{tab:datasets}. The first dataset is TUT Rare Sound Events 2017  \cite{DCASE2017challenge}, which is used in the challenge of Detection and Classification of Acoustic Scenes and Events (DCASE) 2017, as task 2. The second dataset is TAU Spatial Sound Events 2019 - Ambisonic, which is used in the challenge of DCASE 2019 \cite{DBLP:journals/jstsp/AdavannePNV19}, as task 3. 

Both datasets consist of synthetic mixtures created by mixing isolated sound event clips with background sounds. Previous sound event detection studies \cite{DBLP:journals/taslp/CakirPHHV17, DBLP:conf/icassp/ParascandoloHV16} use synthetic datasets as primary evaluation datasets, since the timestamps of sound events in these datasets are precise and consistent. In contrast, real-life recordings use manual annotation, where the subjectivity may lead to inconsistency and possible errors in the labels. The two  datasets in this study are chosen to represent scenarios with different sound event densities, which largely affects the active learning performance.

\begin{table}[ht]
\normalsize
\begin{center}
    \begin{tabular}{l| l| l }
    Dataset & \thead{TUT Rare \\Sound Events \\2017} & \thead{TAU Spatial \\ Sound Events \\2019} \\
    \hline
    Total duration & 25 h & 6 h 40 m\\ 
    Training set duration & 12 h 30 m & 5 h \\ 
    Target event classes & 3 & 11\\ 
    EBR & [-6 db, 0 db, 6 db] & 30 db\\ 
    Recording length & 30 s & 1 m \\
    Events per minute & 1 & 55\\ 
    \end{tabular}
\end{center}
\caption{A Summary of datasets used in the evaluation, explained in Section IV.A.}

\label{tab:datasets}
\end{table}

\subsubsection{TUT Rare Sound Events 2017}
TUT Rare Sound Events 2017 dataset, referred to as rare sound dataset later, is created by mixing isolated target sounds from Freesound with background audio in TUT Acoustic Scenes 2016 dataset \cite{DBLP:conf/eusipco/MesarosHV16}. There are three target event classes: baby cry, gunshot, and glass breaking. Most gunshot and glass breaking sounds are short, lasting around 200 milliseconds. In comparison, baby cry events are longer, typically ranging from one to four seconds. The background consists of sounds from 15 classes of real acoustic scenes, 78 instances each class. The acoustic scenes are bus, cafe/restaurant, car, city center, forest, grocery store, home, lakeside beach, library, metro station, office, residential area, train, tram,  and urban park.

All the background audio tracks last 30 seconds. The sampling rate is 44100 Hz. An audio signal in the rare sound dataset might be either pure background or a target event mixed with a background. The event-to-background ratio (EBR) in dB is randomly chosen from $\{-6, 0, 6\}$, and the positioning of the target sound in a mixture is also random. The sound events are rare in this dataset, on average one event per minute.

The original rare sound dataset is split into a development training set, development test set, and evaluation set. Each split of the dataset contains mixtures created with a separate set of background and target sounds. In this study, the development training set is used for training, and the development test set is used for evaluation. Both the training and test set contains approximately 1500 audio signals, with 250 target events of each class. 

\subsubsection{TAU Spatial Sound Events 2019}
The dataset TAU Spatial Sound Events 2019 dataset, is originally a spatial audio dataset, which is used for sound event detection and spatial localization task in DCASE 2019 challenge. The dataset is synthetic, and the source of the mixtures are sound events from 11 classes, with 20 instances in each class. Each recording in the spatial sound dataset has around one-minute duration, which is mixed with target sound events. On average, each minute of
the signal contains 55 events, randomly positioned, with possibly overlapping in time. The background is relatively quiet and the EBR of the mixtures is about 30 dB. 

The original sampling rate of the dataset is 48 kHz. In the experiments, the recordings are resampled to 44.1 kHz, to match the sampling rate of the pre-trained embedding extraction model. The audio in this dataset has four channels, however, only the first channel is used in this study, since this study does not deal with multi-channel audio.

Similar to the usage of the rare sound dataset, this study uses only the development set, ignoring the evaluation set in the challenge. Four-fold cross-validation is used, following the original setup of the dataset. 

\subsection{Evaluation metric}
In this study, a segment-based error rate (ER) is used to evaluate the performance of a SED model \cite{Mesaros2016_MDPI}. The segment length in the segment-based evaluation is one second, which is a common setup in sound event detection studies, such as DCASE 2017 task 3. 

The aim of active learning is to optimize the accuracy of learned SED models with a limited labeling budget. Thus, the active learning performance is evaluated by ER as a function of the labeling budget, which is given in proportion to the whole training set.

\subsection{Basic experimental setups}

Experiments are made to evaluate each component in the proposed active learning system. This section describes common setups used through all the experiments in the evaluation. 

When computing the spectrogram, the frame length is 40~ms and hop length is 20~ms. In each frame, the signal is windowed with the Hanning window and then log-mel energies in 128 bands are calculated. The gated CNN pre-trained with AudioSet maps a log-mel spectrogram into an embedding sequence with the same number of frames and 256 dimensions. 

The likelihood of change is estimated for each frame based on the past 24 frames and the future 24 frames, aggregating to an analysis window of one second. Detected change points can be closer than one second, for example, the second and third change point in Fig 2. However, annotating very short segments can be difficult in practice. The actual annotation effort is underestimated, when the annotator needs to listen to the extra context of a candidate segment for annotation. In order to avoid very short segments, the change points detected within one second to the previous ones are skipped when generating the candidate segments. As a result, the minimum length of the generated segments is one second.

In the simulation of the labeling process, the ground truth labels are initially hidden to the system. Upon the label query on a segment, annotated labels are simulated according to the ground truth. When a ground-truth sound event overlaps a queried candidate segment with more than 0.1 seconds, a weak label is generated, associating the event class with the segment. It is presumed that an event shorter than 0.1 seconds cannot be perceived by an annotator. 

A SED model is trained with simulated annotations and the performance is benchmarked when the number of simulated labels reaches an evaluated labeling budget. In this study, the following proportions of the training data as labeling budget are evaluated: $1\%, 2\%, 3\%, 4\%, 5\%, 6\%, 7\%, 8\%, 9\%, 10\%, 20\%, 100\%$. During the training of a SED model in each iteration, one-third of the labeled data is randomly chosen for validation. 

The experiments on the TUT Rare Sound dataset are repeated five times, and the average performance is reported. The 4-fold validation experiments on the TAU Spatial sound dataset are repeated twice, and the average of the eight results is reported.

In all the experiments with reported results, the same network architecture is used. A  preliminary study was made to investigate the effect of the model complexity with low labeling budget: we tested using a single GRU layer instead of three when only 1\% of the training data was labeled. As a result, the performances are similar among the tested models with different number of layers.

\subsection{Experiments}

\begin{table*}[h]
    
    \centering
    
    \begin{tabular}{l|l|l|l|l|l}
    & System & Annotation unit & Label type & Sample selection method & Training input\\
    \hline
    \multirow{2}{*}{Experiment A1} & 1 & variable-length segment & weak label & random sampling & \textbf{recordings}\\
    
    & 2 & variable-length segment & weak label & random sampling & \textbf{segments} \\
    
    \hline
    \multirow{2}{*}{Experiment A2} & 3 & \textbf{variable-length segment} & strong label & random sampling & recordings\\
    
    & 4 & \textbf{recording} & strong label & random sampling & recordings\\
    
    \hline
    \multirow{3}{*}{Experiment B} & 1 & variable-length segment & weak label & \textbf{random sampling} & recordings \\
    
    & 5 & variable-length segment & weak label & \textbf{mismatch-first farthest-traversal} & recordings\\
    
    & 6 & variable-length segment & weak label & \textbf{uncertainty sampling} & recordings \\
    
    \hline
    
    \multirow{2}{*}{Experiment C} & 5 & \textbf{variable-length segment} & weak label & mismatch-first farthest-traversal & recordings \\
    & 7 & \textbf{fixed-length segment} & weak label & mismatch-first farthest-traversal & recordings \\
    \end{tabular}
    \caption{A summary of experiments. Bold font is used to highlight the investigated aspect in each experiment.}
    
    \label{tab:experimetal_setup}
\end{table*} 

In order to evaluate each component in the proposed active learning system, four experiments have been made, as is summarized in Table~\ref{tab:experimetal_setup}. 

The proposed system uses variable-length segments as candidate segments for annotation. In order to preserve the context for the annotated segments, the original recordings are used as training inputs, regarded as partially labeled sequences. Experiment A evaluates the effect of preserving the context. Experiment A1 investigates the training input. System 1 uses full recordings as training inputs as is proposed, whereas System 2 uses only annotated segments as training inputs. Experiment A2 investigates the annotation unit. System 3 uses variable-length segments as an annotation unit as is proposed, whereas System 4 uses a full recording as an annotation unit. Strong labels are used in experiment A2 since weak labels are not informative for full recordings in the TAU Spatial Sound dataset, where most recordings include all the 11 sound event classes.  During the model training with strong labels, the attention layer is not used and the training loss is directly computed as the binary cross-entropy between the target and the class probability output on a frame basis. Random sampling is used in all the systems in Experiment A. 

Experiment B focuses on the sample selection method. It compares mismatch-first farthest-traversal, with two reference methods based on random sampling and uncertainty sampling. Random sampling is used in System 1, which is also used in Experiment A1. System 5 uses mismatch-first farthest-traversal, and System 6 uses uncertainty sampling. In random sampling, each candidate segment has an equal probability of being selected. In uncertainty sampling, the certainty of predicting a class $c$ is measured as $ 2\times|o_c - 0.5|$, where $o_c$ is the weak label output or segment-wise class probability. The overall prediction certainty on a sample is defined as the minimum prediction certainty over all the classes. Since uncertainty sampling and mismatch-first farthest-traversal are batch mode active learning, the performance depends on the size of a selection batch. Typically a smaller batch size leads to better accuracy, but it requires more training time. In this experiment, the selection batch size is set to 0.5\% of the whole trained set, which is about 150 segments in the TUT Rare Sound dataset and 60 segments in the TAU Spatial Sound dataset. The batch size is chosen for convenience, since the performance of the learned SED model is reported after every two selection batches, according to the evaluated labeling budget.

Experiment C focuses on the proposed segmentation method based on change point detection. System 5 is a combination of all proposed components in this study. In comparison to System 5, System 7 uses segments with a fixed-length of two seconds. The total number of fixed-length segments is similar to the total number of variable-length segments generated using change point detection.

\subsection{Experimental results}
\begin{figure*}[t]
  \centering
  \includegraphics[trim={40mm 160mm 40mm 20mm},clip,width=0.99\textwidth]{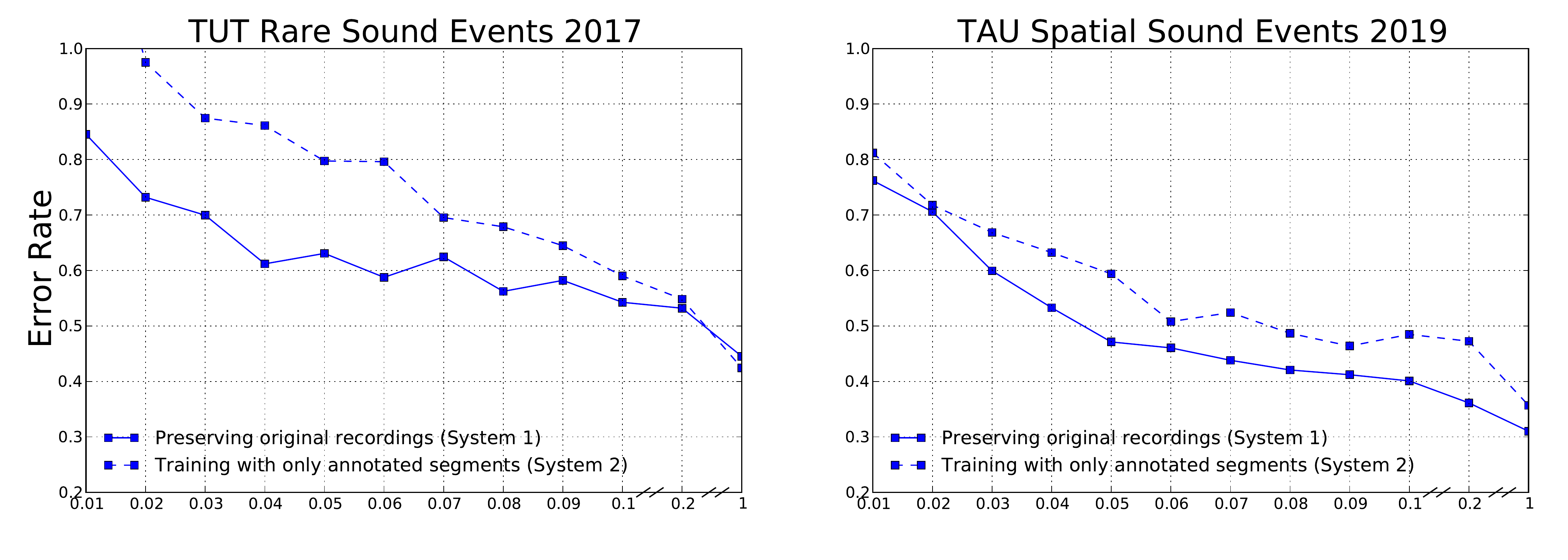}
  \caption{Error rate of learned models as the function of labeling budget for methods that use different training inputs, corresponding to experiment A1. 
  }
  \label{fig:seg_plot1}
  
\end{figure*}

\begin{figure*}[t]
  \centering
  \includegraphics[trim={40mm 160mm 40mm 20mm},clip,width=0.99\textwidth]{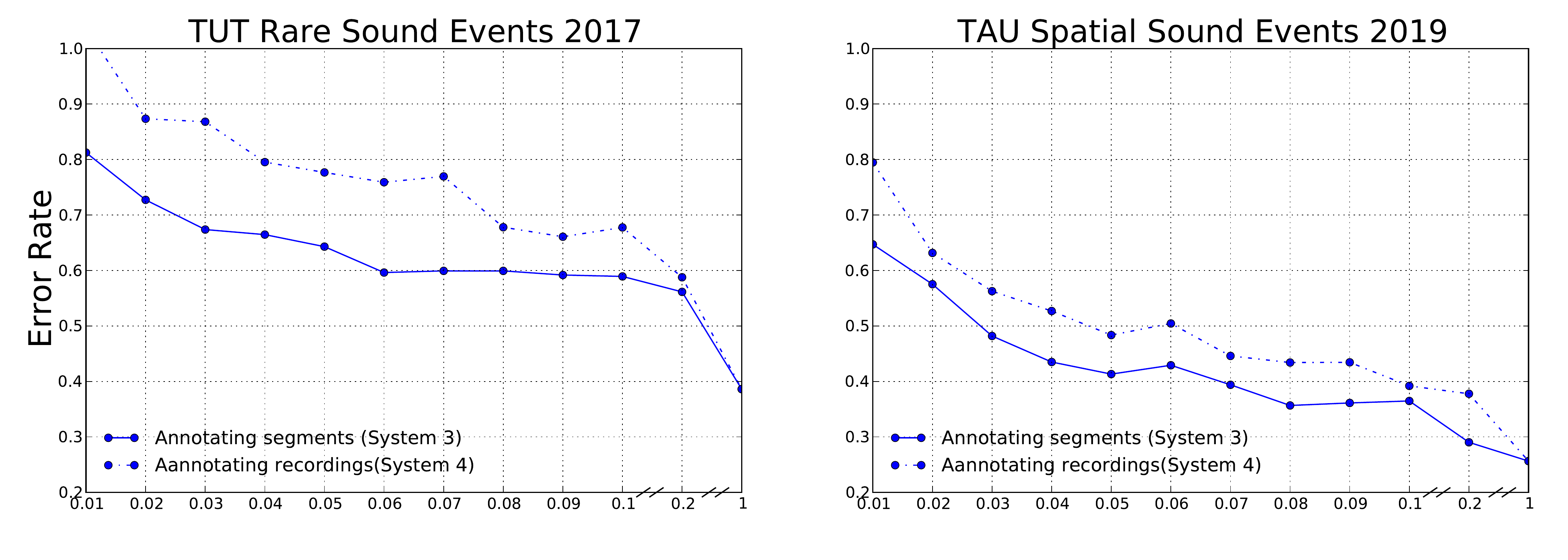}
  \caption{Error rate of learned models as the function of labeling budget for methods that use different annotation units, corresponding to experiment A2.
  }
  \label{fig:seg_plot2}
\end{figure*}

The results of experiment A1, illustrated in Figure~\ref{fig:seg_plot1}, show that preserving original recordings as the context clearly outperforms training with only annotated segments. In some cases, more than 60\% of the labeling budget can be saved to achieve the same accuracy. A sound event is sometimes detected not only based on the audio signal where the event happens but also the difference compared to the background sounds in the temporal context, preserved in the original recordings. The results of experiment A2, illustrated in Figure~\ref{fig:seg_plot2}, show that annotating segments is more efficient compared to annotating full recordings. The segments randomly sampled from all the recordings have typically higher diversity, in comparison to a small amount of fully annotated recordings. In addition, by comparing the results of System 1 and System 3, close performance is achieved by using attention pooling with weak labels, compared to using strong labels.

\begin{figure*}[t]
  \centering
  \includegraphics[trim={40mm 160mm 40mm 20mm},clip,width=0.99\textwidth]{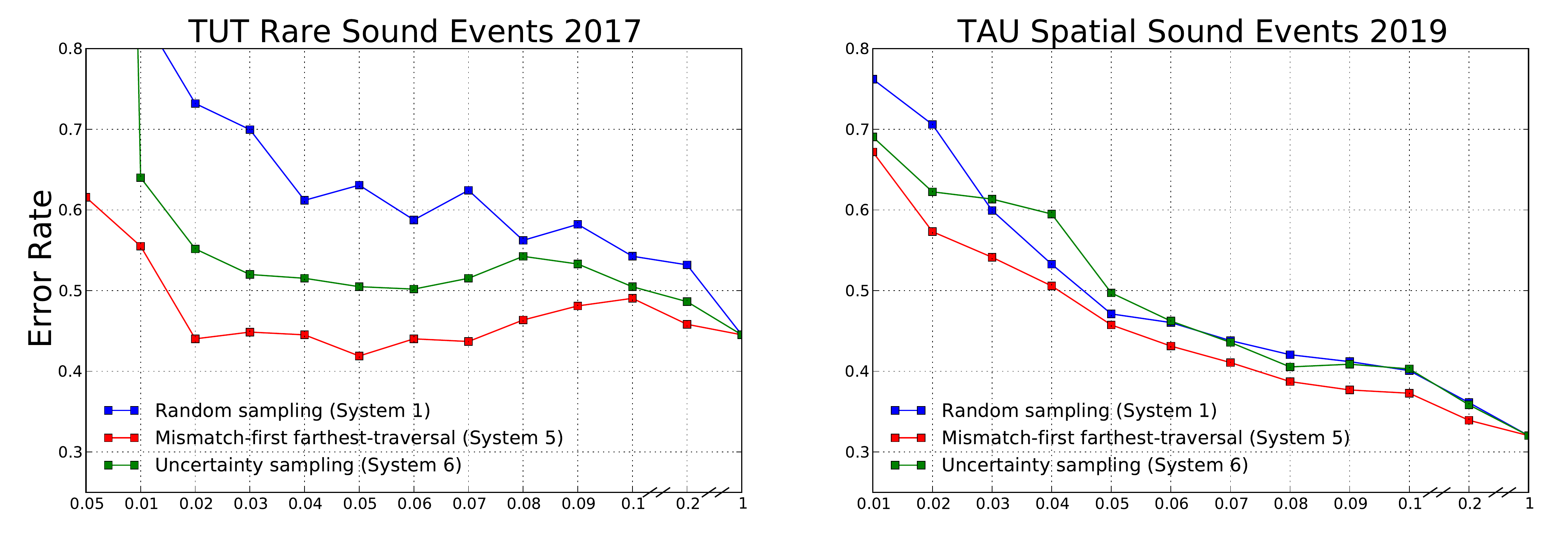}
  \caption{Error rate of learned models as the function of labeling budget for different sampling methods, corresponding to experiment B.  
  }
  \label{fig:query_plot}
\end{figure*}

\begin{figure*}[t]
  \centering
  \includegraphics[trim={40mm 160mm 40mm 20mm},clip,width=0.99\textwidth]{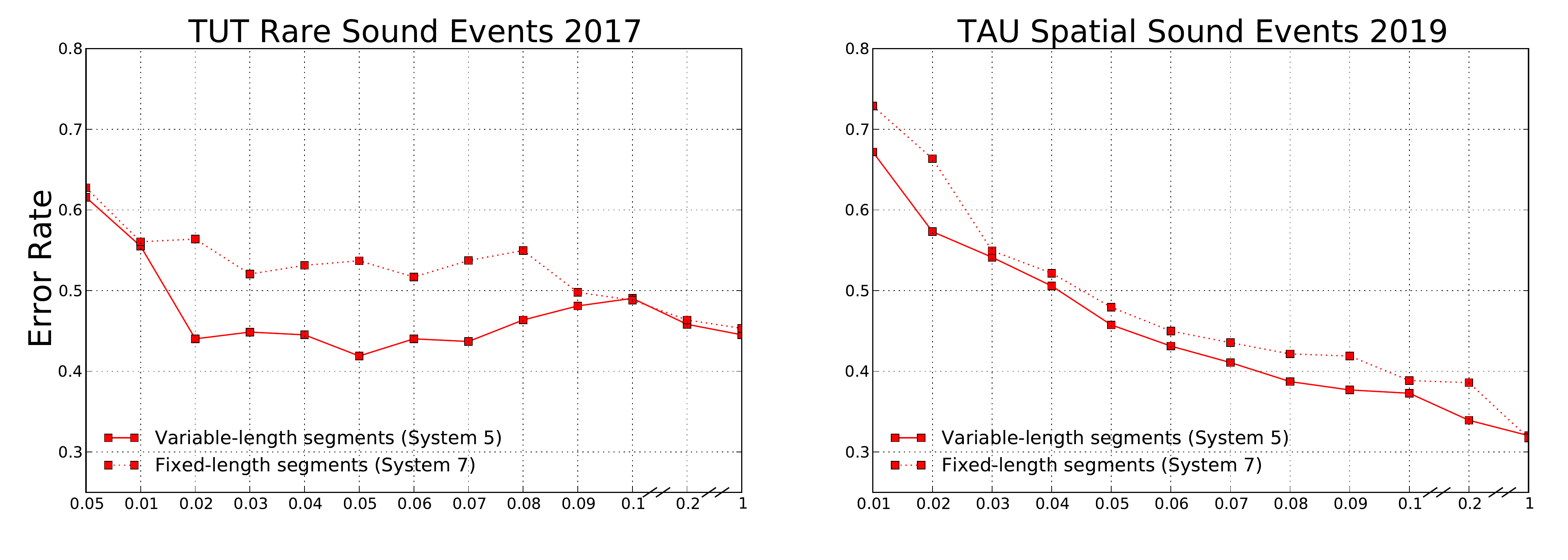}
  \caption{Error rate of learned models as the function of labeling budget for different segmentation methods, corresponding to experiment C.  
  }
  \label{fig:query_plot2}
\end{figure*}

The experimental results comparing the sampling methods are illustrated in Figure~\ref{fig:query_plot}. The results show that the proposed method outperforms reference methods with all evaluated labeling budgets. 

In the experiments on the TUT Rare Sound dataset, the proposed method outperforms reference methods to a large extent. Most of the training data have little relevance to the target problem since the target sound events are rare in this dataset. Therefore, the annotation effort can be greatly reduced by selective sampling, if irrelevant data can be ruled out in the sample selection. In addition, uncertainty sampling also outperforms random sampling to a large extent. 

Remarkably, the proposed active learning method requires only 2\% of the training data to be annotated to achieve similar performance, compared to annotating all the data. Surprisingly, the best performance is achieved by annotating only 5\% of the training set. The sound events are rare in the dataset, and most of the segments containing target events are selected within the first 5\% of the training set. By the time when 5\% of the training data is labeled in a typical case, the segments containing a target event comprise 35\% of the labeled data, whereas, only 1.25\% of the unlabeled data contains a target event. Although more labeled data is available when labeling budget increases, the high label distribution bias has a negative effect on the accuracy of learned models. As a result, the accuracy does not improve with increasing labeling budget.

In the experiments on the TAU Spatial Sound dataset, The proposed method slightly outperforms the two reference methods. In the TAU Spatial Sound dataset, target sound events are dense. In principle, little improvement can be made with selective sampling, when majority of the dataset are relevant to the target SED problem. In this case, the proposed method cannot save much annotation effort.

Combining the effect of sample selection and training with original recordings as context, a clear improvement in performance can be made with the proposed system. This can be evaluated by comparing System 5 with System 2. To achieve ER of 0.55 in the TUT Rare Sound dataset, System 2 requires 20\% of the training set as a labeling budget. In comparison, the proposed method, System 5 requires annotating only 1\% of the training set. To achieve ER of 0.5 in the TAU Spatial Sound dataset, System 2 requires 6\% of the training set as labeling budget. In comparison, System 5 requires annotating only 4\% of the training set.

The experimental results comparing the two segmentation methods are illustrated in Figure~\ref{fig:query_plot2}, when mismatch-first farthest-traversal is used. The experiments show that variable-length segments lead to better performance. Mismatch-first farthest-traversal largely depends on the similarity analysis. Since fixed-length segments often contain part of events, the similarities between fixed-length segments are less relevant to their labels, compared to the similarities between variable-length segments, which is targeted to contain complete events. 

\section{Conclusion}
In this study, we propose an active learning system for sound event detection (SED), which targets on optimizing the accuracy of a learned SED model with limited annotation effort. The proposed system analyzes an initially unlabeled audio dataset, querying for weak labels on selected sound segments from the dataset. A change point detection method is used to generate variable-length audio segments. The segments are selected and presented to an annotator, based on the principle of mismatch-first farthest-traversal. During the training, full recordings are used as input to preserve the long-term context for annotated segments.

Experimental results show that training with original recordings as a context for annotated segments clearly outperforms training with only annotated segments. Mismatch-first farthest-traversal clearly outperforms reference sampling methods based on random sampling and uncertainty sampling. The performance of mismatch-first farthest-traversal depends on the segmentation method that generates the candidate segments. Variable-length segments generated by change point detection lead to clearly better performance than fixed-length segments.

Overall, the proposed method effectively saves annotation effort to achieve the same accuracy, with respect to reference methods. The amount of annotation effort can be saved depends on the distribution of target sound events in the training dataset: a larger amount of annotation effort can be saved when the target sound events are rare. On the dataset with rare events, more than 90\% of labeling budget can be saved by using the proposed system, with respect to a system that uses random sampling and annotated segments only for model learning. Notably, by annotating 2\% of the training data, the proposed method achieves the same accuracy as training with all the data.

In future work, the optimal combination of active learning and semi-supervised learning methods can be studied for SED. Recent semi-supervised learning studies, particularly those based on the mean-teacher method \cite{DBLP:conf/nips/TarvainenV17}, have been shown effective for SED problems in DCASE 2019 task 4 \cite{Turpault2019}. We expect that more annotation effort can be saved, by incorporating semi-supervised learning to further utilize the unlabeld part of the dataset.  

\ifCLASSOPTIONcaptionsoff
  \newpage
\fi

\bibliographystyle{IEEEtran}
\bibliography{mybib}

\begin{IEEEbiography}[{\includegraphics[width=1in,height=1.25in,clip,keepaspectratio]{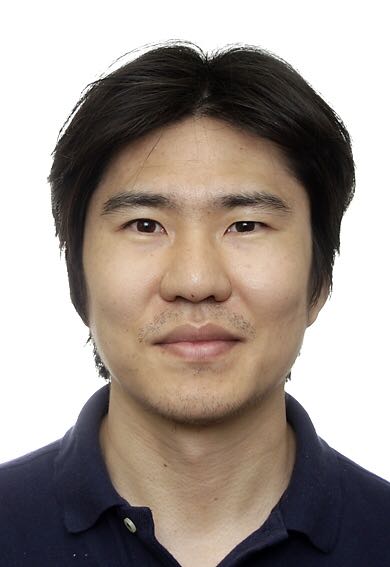}}]{Zhao Shuyang}

received the M.Sc. degree in signal processing from Tampere University of Technology (TUT), 2014. Since 2013, he has been working in Audio Research Group in TUT, where he is currently working towards ph.D. degree. His main research interests include audio content analysis and machine learning.
\end{IEEEbiography}

\begin{IEEEbiography}[{\includegraphics[width=1in,height=1.20in,clip,keepaspectratio]{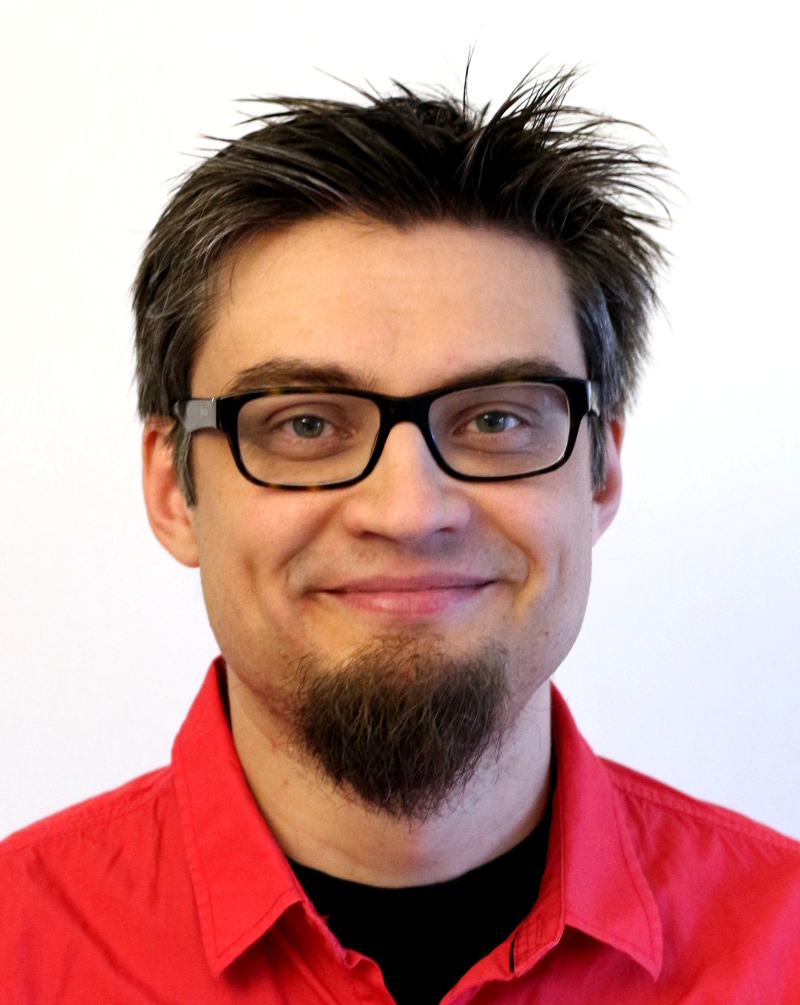}}]{Toni Heittola} is a doctoral student at Tampere University, Finland. He received his M.Sc. degree in Information Technology from Tampere University of Technology (TUT), Finland, in 2004. He is currently pursuing the Ph.D. degree at Tampere University. His main research interests are sound event detection in real-life environments, sound scene classification and audio content analysis.
\end{IEEEbiography} 

\begin{IEEEbiography}[{\includegraphics[width=1in,height=1.20in,clip,keepaspectratio]{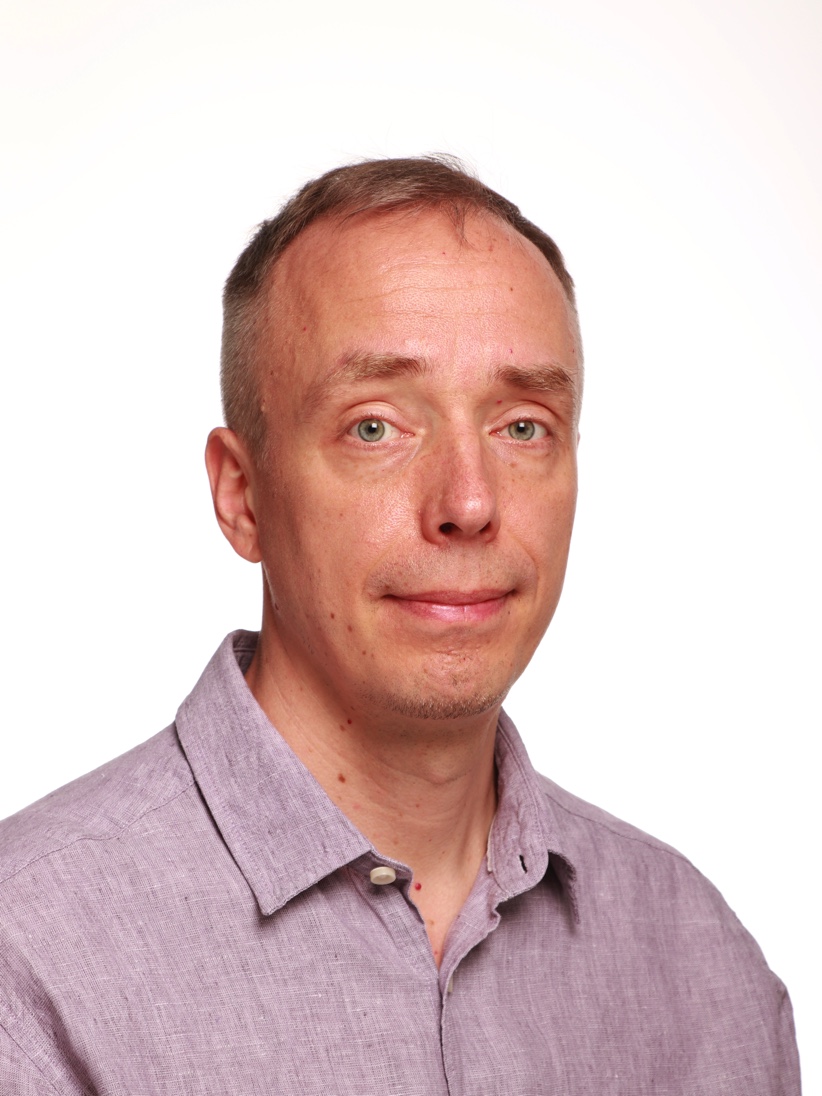}}]{Tuomas Virtanen}
Tuomas Virtanen is Professor at Tampere University, Finland, where he is leading the Audio Research Group. He received the M.Sc. and Doctor of Science degrees in information technology from Tampere University of Technology in 2001 and 2006, respectively. He has also been working as a research associate at Cambridge University
Engineering Department, UK. He is known for his pioneering work on single-channel sound source separation using non-negative matrix
factorization based techniques, and their application to noise-robust speech recognition and music content analysis. Recently he has done significant contributions to sound event detection in everyday environments. In addition to the above topics, his research interests include content analysis of audio signals in general and machine learning. He has authored more than 190 scientific publications on the above topics, which have been cited more than 9000 times. He has received the IEEE Signal Processing Society 2012 best paper award for his article "Monaural Sound Source Separation by Nonnegative Matrix Factorization with Temporal Continuity and Sparseness Criteria" as well as three other best paper awards. He is an IEEE Senior Member, member of the Audio and Acoustic Signal Processing Technical Committee of IEEE Signal Processing Society, and recipient of the ERC 2014 Starting Grant.
\end{IEEEbiography}

\end{document}